# Seasonal and Radial Trends in Saturn's Thermal Plasma Between the Main Rings and Enceladus

M.K. Elrod[1,2], W-L Tseng[3], A. K. Woodson[1], R. E. Johnson[1]

[1] University of Virginia, Charlottesville, VA
[2] National Institute of Aerospace, Hampton, VA
[3] Southwest Research Institute, San Antonio, TX

**Abstract**

A goal of Cassini's extended mission has been to examine the seasonal variations of Saturn's magnetosphere, moons, and rings. Recently we showed that the magnetospheric plasma between the main rings and Enceladus exhibited a time dependence that we attributed to a seasonally variable source of oxygen from the main rings (*Elrod et al.,* 2012). Such a temporal variation was subsequently seen in the energetic ion composition (*Christon et al.,* 2013). Here we include the most recent measurements by the Cassini Plasma Spectrometer (CAPS) in our analysis (*Elrod et al.,* 2012) and modeling (*Tseng et al*., 2013a) of the temporal and radial dependence of the thermal plasma in the region between the main rings and the orbit of Enceladus. Data taken in 2012, well past equinox for which the northern side of the main rings were illuminated, appear consistent with a seasonal variation. Although the thermal plasma in this region comes from two sources, the extended ring atmosphere and the Enceladus torus that have very different radial and temporal trends, the heavy ion density is found to exhibit a steep radial dependence that is similar for all years examined. Using our chemical model, we show that this dependence requires a radial dependence for Enceladus torus than differs from recent models or, more likely, enhanced heavy ion quenching with decreasing distance from the edge of the main rings. We examine the possible physical processes and suggest that the precipitation of the inward diffusing high-energy background radiation onto the edge of the main rings could play an important role.

## 1. Introduction

The thermal plasma in the inner magnetosphere, from just outside the main rings to just inside the orbit of Enceladus (~$2.4R_S – 3.8R_S$; 1 $R_S$ = 1 Saturn Radius = 60,300km), is studied using Cassini Plasma Spectrometer (CAPS) data from 2004 through 2012. The goal of this paper is to further examine the temporal and radial variation in the heavy ion plasma (*Elrod et al.,* 2012). The $O_2^+$ observed by the CAPS instrument over the main rings (*Tokar et al.,* 2006) was suggested to be formed from $O_2$ produced by photo-decomposition of the icy ring particles (*Johnson et al.,* 2006a). A fraction of this $O_2$ is subsequently scattered, forming an extended ring atmosphere, in which the molecules are ionized contributing to the plasma in Saturn's inner magnetosphere (e.g., *Johnson et al.,* 2006a; *Martens et al.,* 2008). Since such a source depends on the illumination of the ring plane, the plasma density was predicted to vary over Saturn's orbit as the ring plane illumination varied from the southern hemisphere through equinox to the northern hemisphere (*Tseng et al.,* 2010; 2013a). Water group ions, labeled here as $W^+$ ($O^+$, $OH^+$, $H_2O^+$, and $H_3O^+$) are also directly formed in this region by ionization of neutrals in the Enceladus torus (*Cassidy et al,* 2010; *Smith et al.* 2010; *Tseng et al.,* 2012). While this source probably does not exhibit a seasonal variation, there is evidence that it fluctuates by up to a factor of four (*Smith et al.,* 2010) and appears to depend on the position of Enceladus in its orbit (*Hedman et al.* 2013).

In *Elrod et al.* (2012) we used CAPS data from SOI to 2010 to confirm that the main rings are an important source of $O_2^+$ and $O^+$ ions inside the orbit of Mimas. We also showed that there was a steep drop in $O_2^+$ density, as well as in the total heavy ion thermal plasma, over that time period. Consistent with this, the Magnetospheric

Imagining Instrument (MIMI) recently reported a decrease from SOI through 2010 in energetic $O_2^+$ density as compared to the $W^+$ density, followed by a recovery in 2011-2012 (*Christon et al., 2013*). More recently a seasonal variation was reported equatorial electron densities measured by the Radio and Plasma Wave Instrument (RPWS) (Persoon et al. 2013a).

The present study of the thermal plasma extends our earlier time frame by including the 2012 data in order to determine whether or not the density increases after equinox. Because seasonal variations were assumed to be due to a ring atmosphere source superimposed on the Enceladus torus source, the seasonal variation might be expected to decrease with distance from the main rings consistent with models of the ring atmosphere. Surprising for all the years studied, the total *heavy ion* density, as well as that for the individual $O_2^+$ and $W^+$ ions, increases relatively steeply with increasing distance from Saturn. This is the case even though the proposed seasonal variation is thought to be primarily due to molecules scattered from the ring atmosphere. After presenting the new results, we consider the implications of these findings.

**2. Analysis**

In order to study the temporal and radial variation in the density and composition of the thermal plasma, we examined passes from SOI in 2004, two in 2007 (doy 162 & 178/179), three in 2010 (doy 062, 154, and 170), and two in 2012 (doy 087 and 105) in the region between ~2.4 $R_S$ – 3.8 $R_S$ (1 $R_S$ = 1 Saturn Radius = 60,300km). All of these passes had a periapsis within the Enceladus orbit at ~4.0$R_S$ and had near equatorial segments which we focus on here. In Fig.1 we show both the orbital paths in the *R* vs. *Z*

81  plane and the local time for each of the passes used in this study. Although there is a

82  significant spread in local time, with the exception of SOI, all the passes were on the

83  dayside, reducing the importance of the suggested night to day variations seen by the

84  Cassini Radio and Plasma Wave Science (RPWS) Langmuir probe (*Holmberg et al.,*

85  2013).

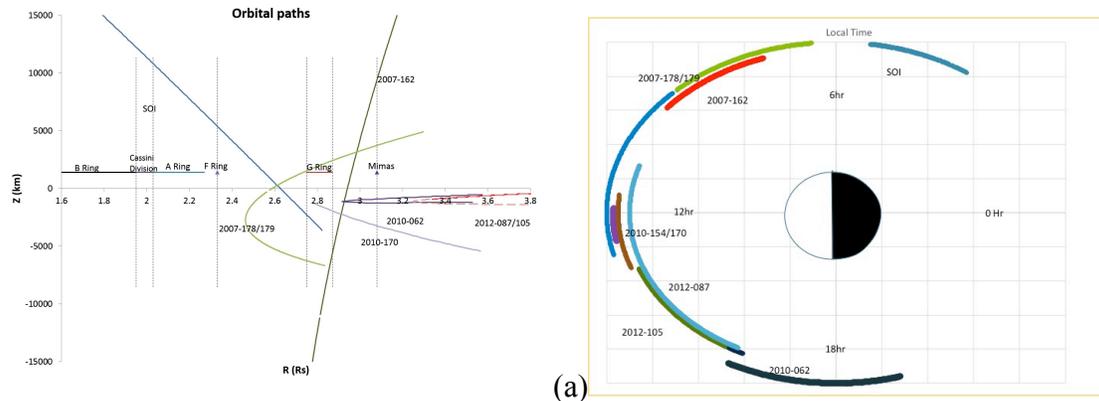

87  (b)

88  Figure 1. Orbits used in this study: a) Position in equatorial radius, $R$ in $R_s$ (1Rs ≈ 60,300 km in the
89  equatorial plane) vs. distance above the equator, $Z$ in km; for consistency in the comparisons and to
90  maximize the count rate, the data presented below comes from the near equatorial segments ($Z$ within
91  ±5,000km). b) Local times during each orbit. With the exception of SOI, most of the data was taken on the
92  dayside between dawn and dusk.

94  In *Elrod et al.* (2012) we described the method of analyzing the CAPS singles

95  data, the correction for the significant background in this region, and the fitting of the

96  measured energy spectrum to obtain the total ion densities as well as the $O_2^+$ and $W^+$

97  fractions. These details are not repeated here. Unlike in *Elrod et al.* (2012) we only

98  analyzed CAPS singles data taken close to the equator, having a minimum number of A-

99  cycles, and at least 400 counts above background in order to reduce the scatter in the

100 data. We also included the 2012 data, which was not available earlier. Although there are

101 variations in the total heavy ion density observed between those passes in Fig. 1

102 occurring in the same year, which we will examine in the future, these variation are
103 typically smaller than the variations between years. Therefore, we show in Fig. 2 the
104 average of the total heavy ion density for each year vs. radial distance from Saturn for
105 those segments close to the equator ($Z < 5000$ km).

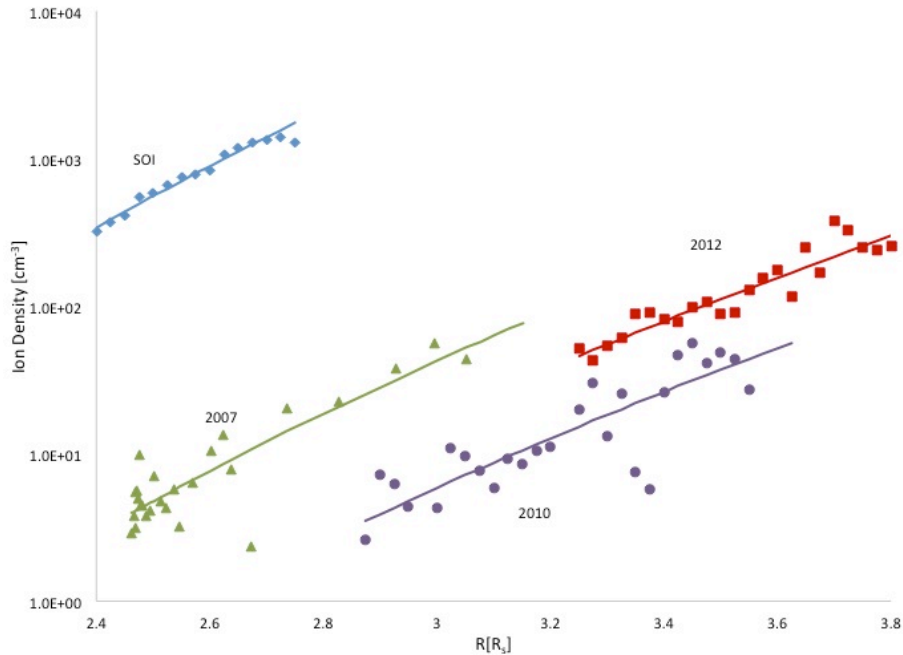

Figure 2. Total heavy ion densities vs. $R$ in $R_s$ from 2004 through 2012 averaged over orbits for each year in Fig. 1a: SOI in 2004 (blue diamonds), 2007 (green triangles), 2010 (purple dots), and 2012 (red squares). Lines are fits: $n_i = c(R/3R_S)^{12}$; where c is the extrapolated value $n_i(3R_S)$: (SOI) blue [$c=4.94 \times 10^3$ /cm$^3$], 2007 green [$c=42.5$/cm$^3$], 2010 purple [$c=5.85$/cm$^3$], and 2012 orange [$c=17.5$/cm$^3$].

113 It is seen in Fig. 2 that there is a clear decrease in the heavy ion density from 2004
114 through 2010 and an increase from 2010 to 2012. Therefore, the lowest densities occur in
115 2010, the orbit closest to equinox (11 August, 2009) with apparent recovery through
116 2012. It is also seen that all densities, regardless of year, increase radially outward. While
117 there are significant changes in the magnitude of the densities between the years, the
118 radial dependence of each of these sets is remarkably similar. In order to better quantify
119 the data we fit a power law in equatorial distance from Saturn, $R$, to each of these

120 averaged data sets. The steep variation is roughly fit using the form $n_i = c\,(R/3R_S)^{12}$.

121 Because the radial range is narrow, powers varying from ~10-14 also give reasonable

122 descriptions. We use a power of 12 in the following and note that the measured radial

123 dependence is relatively steep.

124 As discussed below, the SOI data is dominated by $O_2^+$, likely from the extended

125 ring atmosphere, with temperatures close to the fresh ion pick-up temperature, whereas

126 the later data sets, closer to equinox, are dominated by $W^+$, likely from the Enceladus

127 torus, but with low ion temperatures (*Elrod et al.*, 2012). In spite of the different

128 compositions and temperatures, there is a striking similarity in the radial dependence in

129 all of the years studied.

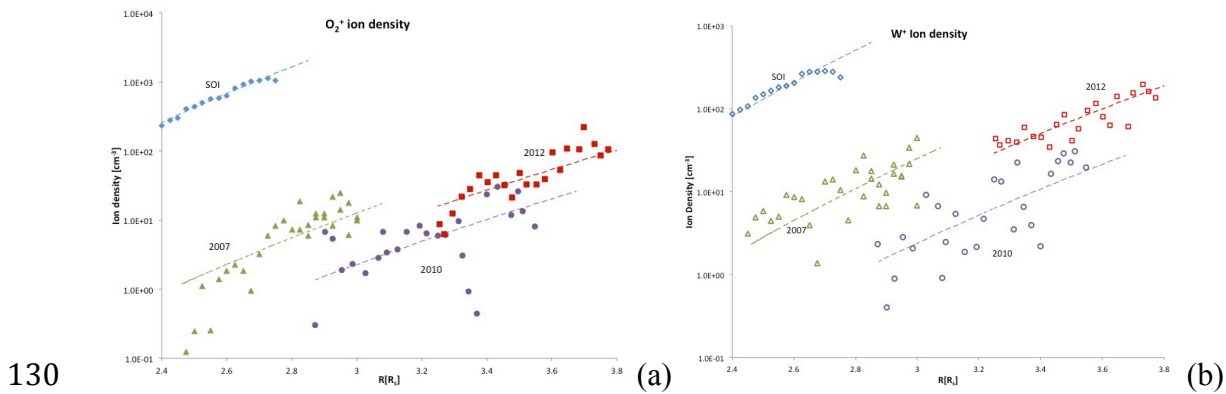

Figure 3. Averaged densities for each year: (a) $O_2^+$, closed symbols; (b) $W^+$ ($O^+$,$OH^+$,$H_2O^+$, $H_3O^+$), open symbols: SOI in 2004 (blue diamonds), 2007(green triangles), 2010 (purple dots), 2012 (red squares). As in Fig. 2 the lines are fits: $n_i(R) = c\,(R/3R_S)^{12}$. The extrapolated values $c \approx n_i(3R_S)$ in cm$^{-3}$ for $O_2^+$ and $W^+$ respectively: (2004: 3.8x10$^3$, 1.2x10$^3$), (2007: 12.8, 25.0), (20010; 2.3, 2.4) and (2012: 6.06, 11.2).

136 Using the analysis in *Elrod et al.* (2012) with the selectivity discussed here, we

137 also show the yearly averaged $O_2^+$ and $W^+$ components of the density in Figs. 3a and b.

138 Because the peaks in the CAPS singles spectra overlap significantly, these data have a

139 much larger scatter, but still exhibit a radial dependence roughly similar to that in Fig. 2.

140 Therefore, we also fit each set using $n_i = c\,(R/3R_S)^{12}$. It is seen to capture the general trend with the exception of the $O_2^+$ 2007 data at the smallest values of *R*. None of these fits are unique, as the $O_2^+$ data can be better fit on average with a slightly smaller power and the $W^+$ with a slightly larger power. Here we use the rough fits in Figs. 2 and 3 to guide the discussion below.

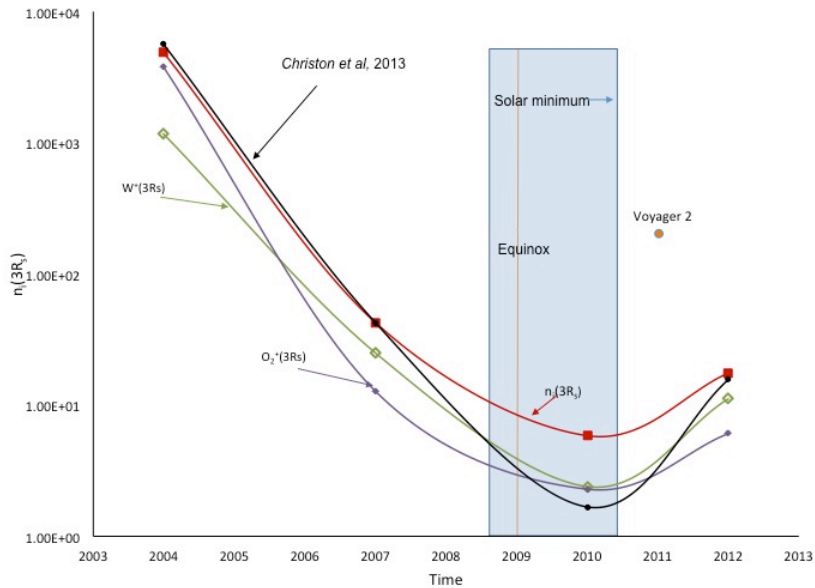

Figure 4. Extrapolated densities at $3R_S$, $n_i(3R_S)$ vs. year, obtained from the value of *c* from the fits in Figs. 2 and 3: (red) total heavy ion density from Fig. 2; (purple) $O_2^+$ and (green)$W^+$ from Fig. 3. Lines are drawn to guide the eye. Using powers of 9.8 and 14.1, which gave slightly better fits to the $O_2^+$ and $W^+$ respectively changed these values by ~ 10%. (Black line) $O_2^+/W^+$ ratio from MIMI data (*Christon et al.*, 2013) times our $W^+$ data. Red dot: Voyager 2 total ion density Aug. 1981, one year after equinox but close to solar maximum. Orange vertical line marks equinox in August of 2009. The grey rectangle marks the region of solar minimum: late 2008 through mid-2010.

In Fig. 4 we plot the fit coefficients, *c*, from Figs. 2 & 3. These are essentially the extrapolated values of the averaged density for each year in the middle of the radial range considered here, $n_i(3R_S)$. Since the ionization rate throughout much of our region is dominated by photo-ionization (e.g., *Sittler et al.*, 2008) we also indicate the rough extent of solar minimum by the grey band and equinox by the vertical line. It is seen that these densities all drop from closest to solstice in 2004 through 2010, closest to equinox and roughly within the solar minimum, and then increase again in 2012. It is also seen that

$O_2^+$ dominates at SOI, whereas $W^+$ dominates in the later years (*Elrod et al.,* 2012) allowing for the significant scatter in the data. A fraction of the $O_2^+$ and $W^+$ produced throughout the magnetosphere gets accelerated to high energies. Therefore, *Christon et al.* (2013) also saw a clear seasonal dependence in the $O_2^+/W^+$ ratio in the MIMI data. Rather than compare their ratio to our densities, we multiply our $W^+$ data by the ratio and display it in Fig. 4. It is seen that the trends are very similar and the post-equinox increase appears to be delayed in both sets of data. This delay is at least in part due to the fact that the northern illumination was predicted to produce a less robust extended ring atmosphere than southern illumination at the same solar inclination angle (*Tseng et al.,* 2010).

Since the recent period of solar minimum occurred close to equinox, the EUV/UV ionization rate also goes through a minimum in the time period indicated in Fig. 4. The photo-ionization rate of $H_2O$ can therefore change by a factor of the order of few (e.g., *Huebner et al.,* 1992). Since solar maximum occurred in about 2000, clearly some of the variation seen in Fig. 4 is due to the changing ionization rate from near the middle of the solar cycle at SOI to a minimum from late 2008 through mid 2010 and then increasing in 2011. Because the EUV/UV photons also determine the photon-induced decomposition of the icy ring particles, the extended the ring atmosphere varies in this time period somewhat faster than predicted in *Tseng et al.* (2010). However, the drop in the thermal plasma density in Fig. 4 is much too large to be accounted for by the change in the solar flux alone. To help separate the seasonal and solar cycle effects, we include the equatorial Voyager 2 plasma density taken about a year after equinox in August 1981, but close to *solar maximum* (*Elrod et al.,* 2012). Although the state of the magnetosphere differed, affecting the electron impact contribution to ionization, and the instrument sensitivities differ, the comparison suggests that the solar ionizing flux is important, but is not the dominant cause of the temporal variations in the thermal plasma seen in Fig. 4.

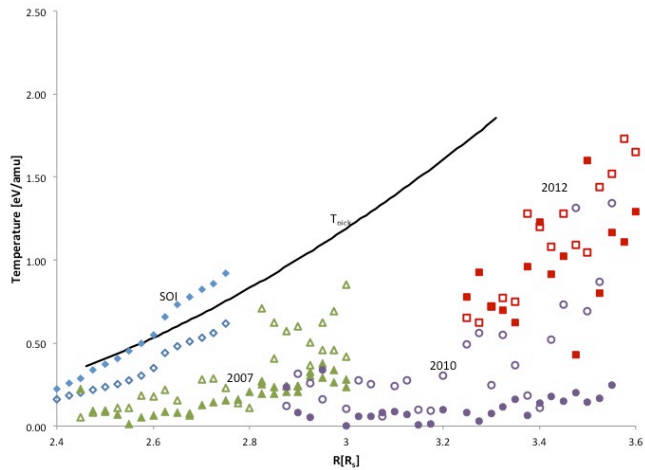

Figure 5. Extracted ion temperatures in (eV/amu) vs. R in $R_S$: for $O_2^+$ and $W^+$ as indicated. Solid line: pick-up temperature/amu for ions freshly formed from a neutral in a circular orbit. The significant scatter in the data is roughly consistent with the uncertainties, not shown for clarity.

In Fig. 5 we show the ion temperatures, $T_i$, for the two species determined from the peak widths. Because the $O_2^+$ and $W^+$ peaks in the CAPS singles data have considerable overlap (*Elrod et al.,* 2012; *Elrod* 2013) there is a significant amount of scatter in these results. However, the trend is clear. At SOI the $O_2^+$ temperatures are close to the pick-up temperature calculated for ions freshly formed from neutrals in roughly circular, Keplerian orbits. This indicates the lifetimes are short compared to collisional cooling, reaction and recombination processes. This is not the case for the other years for which the temperatures for each species are a fraction of the pick-up temperature, a trend that continued in our most recent 2012 data. As proposed in *Tseng et al.* (2013a), this is due to the rapid ion-neutral interactions in this region where the relative ion-neutral speeds are relatively low.

## 3. Discussion of the Results

Earlier we developed a one-box homogeneous ion chemistry model to account for the complex and highly variable plasma environment (i.e. density, composition, and temperature) observed near equator between the main rings and Enceladus (*Tseng et al.*, 2013a). We used models of the Enceladus torus as a source for the water group ions, the scattered ring atmosphere and the Enceladus torus as sources of the light ions, $H^+$ and $H_2^+$, and the scattered ring atmosphere as a source for $O^+$ and $O_2^+$. The density of the Enceladus torus, although varying in its orbit, does not appear to have a seasonal variability (*Smith et al.*, 2010), but the extended ring atmosphere was predicted to vary due to the changing orientation of the ring plane to the flux of solar UV photons as Saturn orbits the sun (*Tseng et al.*, 2010). Here we again use this model and refer the reader to *Tseng et al.* (2013a) for the list of reactions and other details.

Solving a set of chemical rate equations, we showed that the observed temporal variations in the near equatorial *average* in the total heavy ion density and $O_2^+$ density were primarily seasonal due to the predicted variation in the ring atmosphere. The model also *required* an enhancement at SOI caused by a compressed magnetosphere producing an enhanced hot electron component. Due to the low relative ion speeds in this region, the ion–neutral collisions determined both the composition and the surprisingly low ion temperatures found after SOI as seen in Fig. 5. Even though $O_2$ from the extended ring atmosphere is the dominant source of $O_2^+$ at SOI, we also showed that $O_2^+$ is formed by ion-molecule reactions in the magnetosphere, so is present at some level even where the contribution from the extended ring atmosphere is small. Using the dust densities in this region detected by the Cassini Cosmic Dust Analyzer (CDA)(*Kempf et al.*, 2008), the effect of the ion–dust interactions was found to be unimportant.

Our chemical model was applied at an average radial distance, $\sim 3R_S$, and, as stated above, appears to account for the *averaged* plasma densities, composition and temperatures as a function of the season. However, the radial dependence of the ion densities was not examined. Here we show that in going from southern solstice through equinox and onto illumination of the northern side of the ring plane (2004-2012) the radial dependence of these yearly averaged total ion density increases with increasing distance from Saturn as seen in Figs. 2 and 3. Surprisingly, this was found to be the case whether the plasma was dominated by $O_2^+$, as at SOI, or by $W^+$. This dependence is much steeper than the radial dependence of the neutral density in the Enceladus torus (*Cassidy and Johnson*, 2010) and is *opposite* to the radial dependence of the extended ring atmosphere (*Tseng et al.*, 2010). Using even a simple model of reaction pathways, which includes a diffusion time, and the neutral sources, the resulting ion density at equinox varies only slowly with $R$ when using the Enceladus torus source in *Cassidy and Johnson* (2010). Since the average ion densities were about right, either the radial dependence of neutral sources are significantly in error or a quenching mechanism that becomes more important with decreasing distance from the main rings is absent or incorrect.

The difference in the spatial morphology of the source rate and the heavy ion density is especially evident in the SOI data when the plasma is heavily dominated by $O_2^+$ with temperatures close to the pick-up energy. The latter indicates, as stated above, that these ions, formed from oxygen in the extended ring atmosphere, have short lifetimes. The neutral oxygen source clearly decreases with distance from the main rings.

249  However, it is seen in Fig. 3a that the $O_2^+$ density increases with increasing $R$ over the
250  narrow range of $R$ for which there is data. Below we consider those processes that might
251  produce the observed radial dependences from 2004 to 2012.

## 4. Radial Dependence

253  We focus here on the radial variation in the near equatorial densities in Fig. 2.
254  Although the heavy ion data in 2010, nearest to equinox varies between ~4 to ~40/cm$^3$ in
255  going from ~2.9 to ~3.5$R_s$, the ion-chemistry model in *Tseng et al.* (2013a) gives heavy
256  ion densities that are nearly independent of $R$ (~25/cm$^3$ with a slight minimum at 2.75$R_s$).
257  Below we discuss the possible reasons for the difference between the model and
258  measured radial dependence. Before proceeding, we note that heavy ion density in a flux
259  tube (density vertically integrated along a field line) at each $R$ is more closely related to
260  the ionization source rate than is the equatorial density. It can be roughly estimated as the
261  equatorial density times the scale height which is proportional to $\sim T_i^{1/2}$. Because the ion
262  temperature in Fig. 5 increases with increasing $R$, the flux tube content would exhibit an
263  even steeper radial dependence than the near equatorial densities in Fig. 2. Therefore, the
264  difference in the morphology of the heavy ion flux tube content and the spatial
265  morphology of the sources would be even more dramatic.
266  In *Tseng et al.* (2013a) the averaged ion-molecule collisions and ion-electron
267  recombination processes had timescales near equinox of a ~ 10$^5$s throughout this region
268  and ~ 10$^4$s at SOI inside of ~2.8$R_S$. Therefore, processes that might account for the steep
269  decay with decreasing $R$ must have shorter time scales. For example, using the fit to the
270  2010 data in Fig. 2, a quenching process that accounts for the observed radial dependence
271  can have a longer time scale, greater than a few times 10$^5$s at ~ 3.5$R_S$. However, at ~2.5$R_S$

our data would require a time scale ~ 5 x$10^4$s. That is, to account for the steep radial dependence, the lifetime of the heavy ions produced must be smallest closest to the edge of the rings. Although we do not determine the exact quenching mechanism, we examine a number of possible processes: enhanced radial diffusion with ions quenching rapidly on the edge of the main rings; the presence of an increased density of small charged grains on which the ions neutralize; enhanced H or $H_2$ emanating from Saturn's atmosphere or the ring atmosphere increasing the light ion density and quenching the heavy ions; an increased electron-ion recombination rate closer to the rings due to larger electron density and/or smaller electron temperatures.

### *Quenching by Diffusion*

The rate of plasma diffusion in the region of interest has not been examined in any detail. In *Tseng et al.* (2013a) we used a diffusion model consistent with data taken primarily outside the orbit of Enceladus (e.g., *Rymer et al., 2008; Sittler et al.,* 2008). Extrapolating this model into our region, we used an average diffusion time of ~5x$10^6$s. Such times are much too long to affect the radial dependence in the ion density extracted here. Further, due to the predicted radial dependence near the magnetic equator ($\sim R^{-3}$), larger times occur at the smallest $R$, opposite to what would be required to explain our results. Finally, we also note that the ion temperatures vs. $R$ in Fig. 5, particularly for 2007, 2010, and 2012 data, are inconsistent with rapid inward diffusion.

Of course, the standard diffusion model might not be applicable in the region of interest. That is, there are significant and day/night asymmetries observed in the inner magnetosphere that are suggestive of an unexplained noon to midnight electric field (*Homlberg et al.,* 2013) or may related to the orbital varying Enceladus source. Although the observations in the region of interest are sparse, variations in the ion speeds are comparable to the co-rotation speed and could enhance the effective diffusion rate. For instance, *Farrell et al.* (2008) interpreted their 'down-drifting z-mode tones' over the A-ring, measured by the Cassini RPWS instrument at SOI, as being due to rapid unloading of the plasma onto particles in the A-ring. They presumed the plasma diffused inward from the Enceladus torus source and quenched near the edge of the main rings at a rate of ~ 40 kg/s. This would be a significant fraction of the total ion production in the torus inside the orbit of Enceladus. This, in turn, would require diffusion times much more rapid than those discussed above in order to compete with the electron-ion recombination. Since these measurements were *only* available at SOI when the ring atmosphere was maximum and the magnetosphere was active, their observations are probably not applicable to others years. Rather, they are more likely consistent with the quenching of the relatively dense plasma formed just outside of the main rings from the extended ring atmosphere, which dominates the Enceladus torus source in the region of interest at SOI (*Elrod et al.,* 2012).

***Quenching on Small Grains***

Using the dust density observed by the CDA instrument, the ion-dust interaction did not significantly affect the ion loss or cooling in this region (*Tseng et al.,* 2013a). However, there is likely a significant population of submicron-sized dust under the detection limit of this instrument (> 0.9μm). As seen in Fig.1 the region inside $3R_S$, contains the G-ring, the F-ring and the edge of the main rings in which collisions of small icy bodies produce debris (*Tiscareno et al.,* 2013; *Attree et al.,* 2013). This debris can deplete the energetic ions and electrons (*Paranicas et al.,* 2008; *Cuzzi and Burns*, 1988)

318 as also seen in the CAPS background radiation (*Elrod,* 2012). Such debris, in the form of
319 small grains, would be a sink for the thermal plasma with the required radial dependence
320 (highest density at $2.4R_S$ and lowest at $3.8R_S$) suggestive of a ring source of small grains
321 rather than an Enceladus source. Modeling indicates that the resulting dust/grain
322 population typically follows a steep size distribution varying as $1/r_g^a$ with $a \sim$ 4–5 where
323 $r_g$ is the grain radius (*Kempf et al.,* 2008). The ion-dust cross section can be written as [$\pi$
324 $r_g^2 (1 - U_g/E)$] where $U_g$ is the grain potential, which negative in this region (*Jurac et al.,*
325 1996). Here $E$ is either the thermal energy or the relative flow energy between the ion and
326 the grain, depending on whether the grain has been accelerated to corotation or is in a
327 Keplerian orbit with a speed relative to the corotating plasma of $\sim 1.5 \times 10^6$ cm s$^{-1}$ at $2.5R_S$.
328 Presuming impacting ions neutralize on these grains, using an intermediate size
329 dependence, $\sim 1/r_d^{4.5}$, and the relative speed above, with a density of 1$\mu m$ grains of $\sim$
330 $2 \times 10^{-9}$/cm$^3$, the density at $2.5R_S$ would require a contribution from grains with radii down
331 to $\sim$ *10nm*. However, the resulting density of small grains is large enough to significantly
332 deplete the electrons, so that charged grains would dominate the total negative charge in
333 this region. Evidence for this has not been seen, although if the small grain source was
334 highly variable, it might contribute to the significant variability in the electron density in
335 this region reported in *Persoon et al.* (2013a).

336 ***Quenching by Hydrogen from the Main Rings or Saturn***

337 In addition to the possible presence of small grains from the main and tenuous
338 rings, the region outside of the main rings has significant levels of neutrals that decrease
339 in density with increasing *R*. These are atoms and molecules from the icy ring particles
340 (e.g., O, O$_2$ and H, H$_2$; *Johnson et al.,* 2006a; *Tseng et al.,* 2013b) or H from Saturn

341  (*Shemansky et al.,* 2009; *Melin et al.,* 2009; *Tseng et al.*, 2013b). The light neutrals can
342  react with the heavy ions changing the ion composition. However, with the exception of
343  the important reaction, $H + O^+ \rightarrow H^+ + O$, these reactions do not result in a change in the
344  *total* heavy ion density. Therefore, we find that an increased H population, which is also
345  less readily ionized, does not appear to have a major effect on the ion density and
346  composition (*Tseng et al.,* 2103a).
347      The presence of additional neutrals can also act to cool the electrons, especially if
348  they are molecular. In *Tseng et al.* (2013a) we calculate electron density and temperature,
349  $T_e$, self-consistently with the production of and cooling of the ions. That is, we allow the
350  photoelectrons produced to interact and cool by collisions with the molecules, hot
351  electrons, and ions. Since the molecular ion neutralization rate increases with increasing
352  electron density and with decreasing $T_e$, as $\sim T_e^{-1/2}$, the presence of additional neutrals as
353  both a source of additional plasma and cooling of the electrons would cause the
354  recombination rate to increase with decreasing $R$. If the enhanced cooling is due to
355  molecules from the main rings produced by solar UV decomposition of ice particles, it
356  would also be seasonal. That is, it would not be consistent with the dependence seen in
357  Fig. 2: i.e., rapid quenching with decreasing $R$ in 2007 when the seasonal ring atmosphere
358  is predicted to be small. It would require a separate source from the edge of the rings,
359  such as the inward diffusing high-energy background radiation.

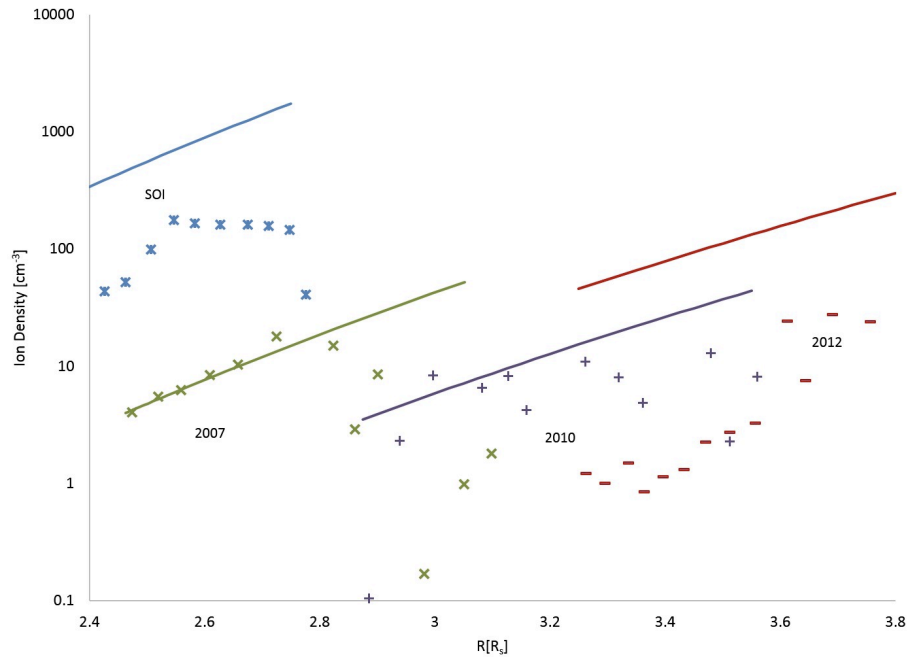

Figure 6. Average ion density vs. $R$. Symbols (x,*, +,-) are the rough estimates of the total light ion ($H^+$, $H_2^+$, $H_3^+$) density, as discussed in the text: colors as in Fig. 2. For comparison, lines are the fits to the heavy ion densities from Fig. 2.

### *Enhanced Light Ion Density*

As discussed above, the light neutrals can result in the production of light ions, $H^+$, $H_2^+$ and $H_3^+$. These are not included in the total heavy ion densities so that the radial dependence of the total ion density might differ significantly from that for the heavy ions. In Fig. 6 we use the CAPS data to extract *very rough* estimates of the light ion densities and overlay them onto the fits for heavy ion densities in Fig. 2. Because the light ion signal usually does not rise above the large background radiation in this region (*Elrod* 2012), these are rough estimates in which a 'feature' in the CAPS data is used when it is seen, but are primarily estimated from the size of the background in the expected location of the peak. Therefore, these results are only for guiding the discussion below.

In spite of the significant scatter, it is seen that the light ions apparently *can* contribute significantly at the smaller values of $R$ in Fig. 5. This is consistent with what we find using the model in *Tseng et al.* (2013a). In our chemical model, $H_2$ from the ring atmosphere plays a significant role at SOI, primarily producing, $H^+$ and $H_2^+$, that account for >50% of the total plasma from 2.5-3.0$R_S$, larger than our rough estimates at SOI in Fig. 6. The light ion fraction decreases at larger $R$, roughly consistent with the trend in Fig. 6. Including the light ion contribution produced by $H_2$ from the ring atmosphere, our modeled rates still can not account for the heavy ion radial dependence measured at SOI.

Because of the uncertainties in the CAPS light ion densities, accurate measurements of the density and temperatures of the electrons in this region are critical. *Persoon et al.* (2013a,b) recently summarized the RPWS electron data, which they show is highly variable. Averaging over a number of Cassini orbits, their fit to that data inside the orbit of Enceladus suggests a much more slowly varying radial dependence (~$R^4$; *Persoon et al.* 2013b) than that seen in Fig. 2 in the heavy ion data. However, as shown in *Elrod et al.* (2012) their densities at SOI are not inconsistent with our ion densities, and there is also rough agreement between their electron densities and the heavy ion densities in Fig. 2 at ~ *3.5$R_S$*. This still allows that the electron densities at ~*2.5$R_S$* could be significantly greater than our measured heavy ion density for passes other than SOI. This would affect the recombination rates and suggest the presence of a significant component of light ions.

The 2007 data is interesting in that Cassini came reasonably close to the main rings at a time well past solstice when the contribution of $O_2^+$ from the ring atmosphere has been predicted to drop significantly (*Tseng et al.,* 2010). It is seen in Fig. 2 that there is significant scatter in the heavy ion data close to the ring, as the densities on the incoming and outgoing passes differ. It is also seen that the light ion densities in 2007 can contribute significantly to the total ion density at the smallest values of $R$. RPWS electron data in 2007 soon to be published (*A. Persoon,* personal communication) suggest that inside ~ 2.9Rs, there is an upturn in the electron density near those values of $R$ closest to Saturn where our light ion estimates appear to contribute significantly. As discussed above, such an upturn would be consistent with a source of light ions produced by coming from inside the region examined.

## 5. Summary

Including new data from post-equinox orbits, selecting only ion energy spectra which had significant counts above background, and averaging over passes occurring in the same year, we confirm that the near equatorial heavy ion density exhibits a significant temporal variation between 2004 and 2012 with a minimum near 2010. Although the ionization rate and photo-induced decomposition rate change during the solar cycle, the large variation in density and composition from SOI to equinox a seasonal effect appears to dominate. That is, the densities are very high and dominated by $O_2^+$ nearest to southern solstice, are smallest close to equinox, and begin to increase post equinox, as seen from the fit parameter to the total ion density in Fig. 4. Although there will be no new data from the CAPS instrument to follow the growth to northern solstice, this interpretation appears to be supported by MIMI observations of energetic heavy ions (*Christon et al.,* 2013) and most recently by the electron data (*Persoon et al.* 2013a). However, with the limited number of orbits in this region that have good data, we allow that other interpretations are possible and the relative importance of the ring atmosphere and solar activity needs to be explored further.

Surprisingly we also show that the heavy ion density exhibits a steep radial dependence over this narrow region (2.4 to 3.8$R_S$), which appears to be roughly independent of the year. Our ion-neutral chemical model (*Tseng et al.,* 2013a) was able to describe the observed temporal dependence in the *average* density, composition and temperatures in this region. In this we used a seasonal ring atmosphere source (*Tseng et al.,* 2010) and Enceladus torus source (*Cassidy and Johnson,* 2010). This model also accounted for the surprisingly low ion temperatures found after SOI, seen again in the recently analyzed 2012 data set (Fig. 5). However, this model could not describe the observed radial dependence of the heavy ion density in Fig. 2, even allowing for the considerable uncertainties.

In the absence of the SOI data set, which is dominated by $O_2^+$ supplied primarily by the ring atmosphere, one might conclude that the *average* densities in the other years are consistent with an Enceladus neutral torus source that is variable within the range suggested by *Smith et al.* (2010). In this case SOI would be exceptional, due to a robust ring atmosphere, with the contribution of the extended ring atmosphere negligible in 2007-2012. If that is the case, then the radial dependence in the heavy ion density seen in those years might suggest that the models of the Enceladus neutral torus should decay more steeply than predicted with decreasing $R$. Since the distribution of neutrals in the torus depends nonlinearly on the density due to neutral-neutral collisions (*Cassidy and Johnson,* 2010; *Cassidy et al.,* 2011), if the densities are lower than those used the radial dependence would be steeper, as seen in models that neglect the neutral-neutral collisions (*Johnson et al.,* 2006; *Smith et al.,* 2010). Recently the variability seen by *Smith et al.* (2010) has been shown to be associated with the position of Enceladus in its orbit

(*Hedman et al.* 2013), so that the model of the neutral torus used in our ion chemistry model needs updating. Such work is now in progress and includes more detail, accounting for interactions with ice grains, the small moons, and the F & G rings within this region.

It is clear, however, that in spite of the large differences in the magnitude of the ion densities in going from 2004 to 2012, there is a similar radial trend over the narrow region (2.4 to 3.8 $R_S$). We point out that this cannot be due to rapid inward diffusion of ions formed from neutrals in the Enceladus torus, which *Farrell et al.* (2008) proposed to explain their SOI data. In the absence of good measurements of the light ion and electron densities for all passes, we suggest this dependence is primarily due to enhanced quenching of the heavy ion plasma with decreasing distance from the edge of the A-ring. Unless the ion diffusion rate in this region differs significantly from estimates at larger $R$, the observed dependence is likely due to material emanating from smaller radial distances. Of particular interest is $H_2$, which plays an important role in our chemical model (*Tseng et al.* 2013a). A likely source is the significant, high-energy, background radiation, which diffuses inward and quenches on the particles in the tenuous rings and at the edge of the A-ring leading to production of $H_2$ from the ice particles.

Since there will be no new CAPS data, understanding this neglected, but extremely interesting, region of the Saturnian system will require more detailed modeling of the extended ring atmosphere, the fate of the ring plasma, the role of the tenuous rings, the quenching of the high energy background radiation, and the variability in the Enceladus source in this region.

## 6. Acknowledgements

We acknowledge support through Southwest Research Institute from a grant for the Cassini Mission from JPL and by a grant from NASA's Planetary Atmospheres program. This work was also supported in part by National Aeronautics and Space Administration, Langley Research Center, under research cooperative agreement No. NNL09AA00A awarded to the National Institute of Aerospace.